\documentclass[12pt,twoside]{article}
\usepackage{xrb2007,graphicx,amssymb,rotating} 
\def\cmmoinsdeux{\mbox{ cm}^{-2}}

\def\microns{\mbox{ } \mu \mbox{m}}

\def\Msol{\mbox{ }M_{\odot}}
\def\Rsol{\mbox{ }R_{\odot}}

\def\mags{\mbox{ magnitudes}}

\def\adeg{^{\circ}}

\def\amin{^\prime}

\def\nh{N_{\rm H}}

\def\ltsima{\; \buildrel < \over \sim \;}
\def\simlt{\lower.5ex\hbox{\ltsima}}            
\def\gtsima{\; \buildrel > \over \sim \;}
\def\simgt{\lower.5ex\hbox{\gtsima}}            

\begin{document}

\session{Obscured XRBs and INTEGRAL Sources}

\shortauthor{Chaty}
\shorttitle{Infrared Observations of {\it INTEGRAL} Sources}

\title{Obscured High Mass X-Ray Binaries and Supergiant Fast X-ray
  Transients: Infrared Observations of {\it INTEGRAL} Sources
\footnotetext{Based on
observations collected at the European Southern Observatory, Chile
(proposals ESO N$\adeg$ 070.D-0340, 071.D-0073, 073.D-0339, 075.D-0773
and 077.D-0721).}}
\author{S. Chaty} \affil{Laboratoire AIM, CEA/DSM - CNRS - Universit\'e
  Paris Diderot, IRFU/Service d'Astrophysique, B\^at. 709, CEA-Saclay,
  FR-91191 Gif-sur-Yvette Cedex, France}

\begin{abstract}
  A new type of high-energy binary system has been revealed by the
  {\it INTEGRAL} satellite.  These sources are being
  unveiled by means of multi-wavelength optical, near- and
  mid-infrared observations. Among these sources, two distinct classes
  are appearing: the first one is constituted of intrinsically
  obscured high-energy sources, of which IGR~J16318-4848 seems to be
  the most extreme example. The second one is populated by the
  so-called supergiant fast X-ray transients, with IGR~J17544-2619
  being the archetype. We report here on multi-wavelength optical to
  mid-infrared observations of a sample constituted of 21 {\it INTEGRAL}
  sources. We show that in the case of the obscured sources our
  observations suggest the presence of absorbing material (dust and/or
  cold gas) enshrouding the whole binary system.  We finally discuss
  the nature of these two different types of sources, in the context
  of high energy binary systems.
\end{abstract}

\vspace{-5mm}
\section{Introduction}

The {\it INTEGRAL} observatory has performed a detailed survey of the
Galactic plane.  The ISGRI detector on the IBIS imager has
discovered many new high energy sources, most of which have been reported in
\cite{bird:2007}\footnote{See also {\em
  http://isdc.unige.ch/$\sim$rodrigue/html/igrsources\-.html}}.  The
most important result of {\it INTEGRAL} to date is the discovery of
many new high energy sources -- concentrated in the Galactic plane, and
in the Norma arm (see e.g. \citeauthor{chaty:2005a}
\citeyear{chaty:2005a}) -- exhibiting common characteristics which
previously had rarely been seen. Many of them are high mass X-ray
binaries (HMXBs) hosting a neutron star orbiting around an O/B
companion, in most cases a supergiant star. They divide into two
classes: some of the new sources are very obscured, exhibiting a
huge intrinsic and local extinction,
and the others are HMXBs hosting a
supergiant star and exhibiting fast and transient outbursts -- an unusual
characteristic among HMXBs.  These are therefore called Supergiant Fast
X-ray Transients (SFXTs, 
\citeauthor{negueruela:2006a} \citeyear{negueruela:2006a};
\citeauthor{sguera:2005} \citeyear{sguera:2005}).
%
High-energy observations are not sufficient to reveal the nature of
the newly discovered sources, since the {\it INTEGRAL} localisation
($\sim 2\amin$) is not accurate enough to unambiguously pinpoint the
source at other wavelengths. Once X-ray satellites such as {\it
  XMM-Newton}, {\it Chandra}, or {\it Swift} provide an arcsecond
position, the hunt for the optical counterpart of the source is open.
However, the high level of absorption towards the galactic plane makes
the near-infrared (NIR) domain more efficient for identifying these
sources.  We first report on multi-wavelength observations of two
sources, one belonging to each class described above, and we then give general
results on {\it INTEGRAL} sources, before discussing them and
concluding.

\section{Observations and Results} \label{observations}

\begin{figure}
  \includegraphics[height=.37\textheight,angle=-90]{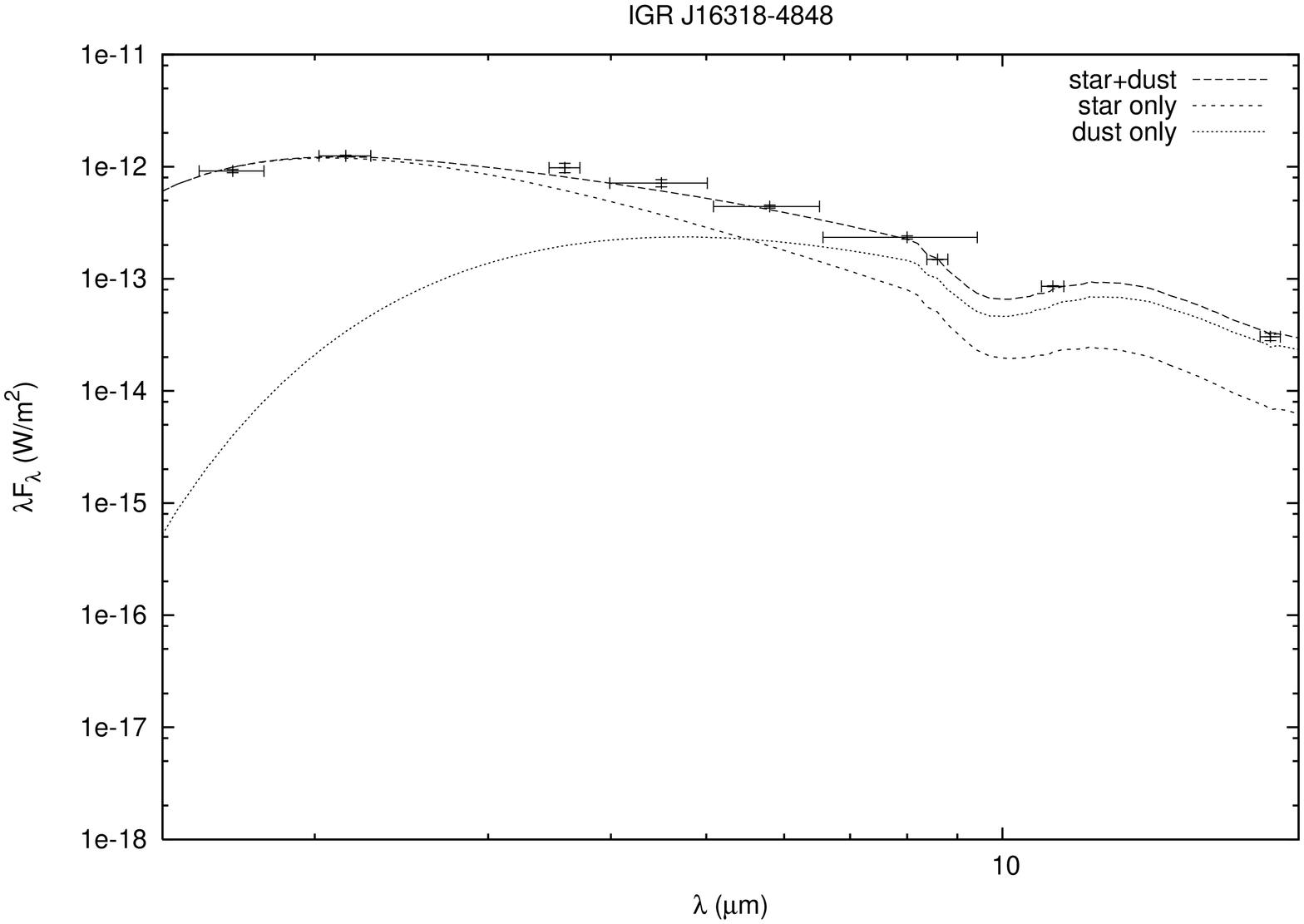}
  \includegraphics[height=.37\textheight,angle=-90]{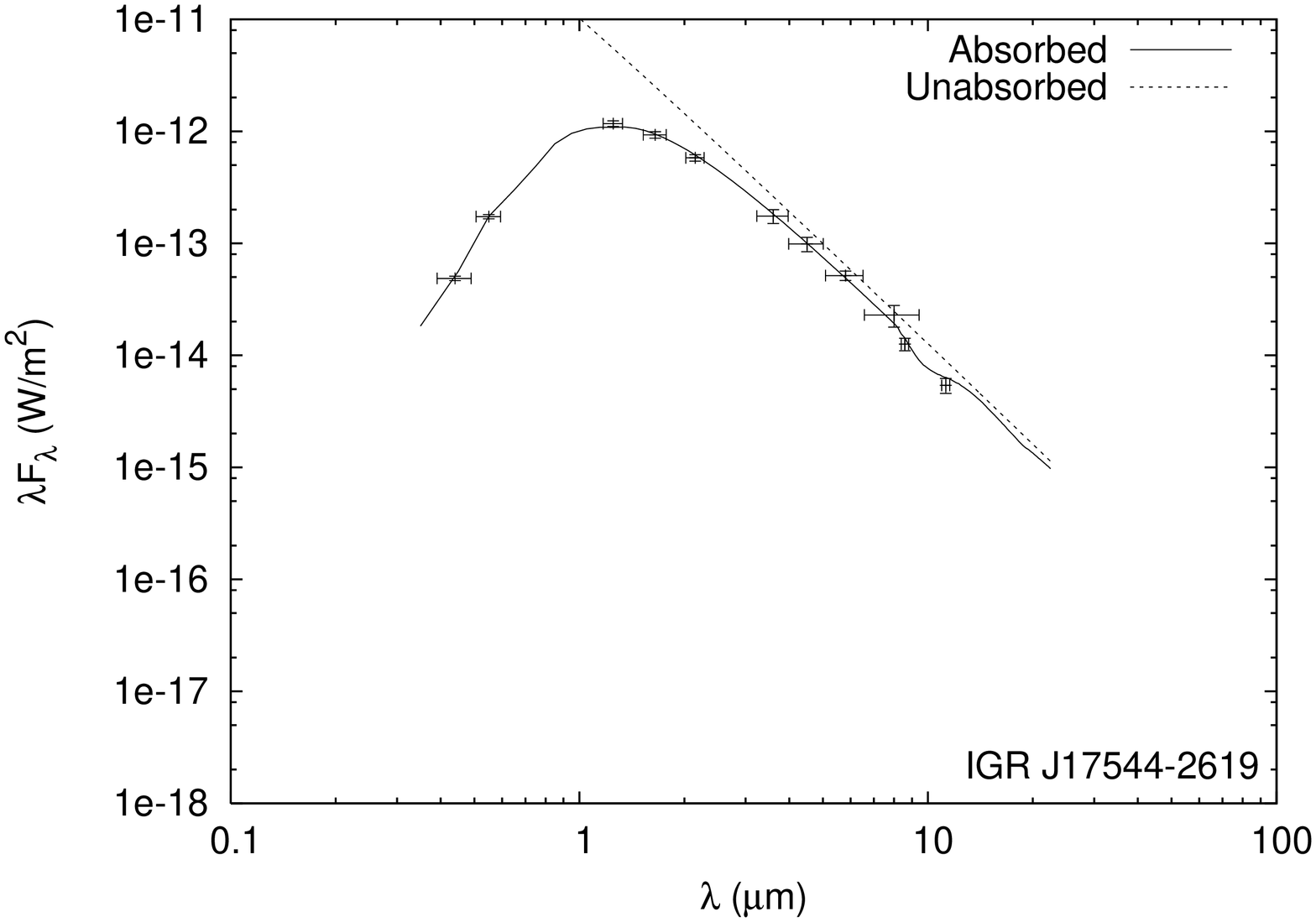}
  \caption{\label{figure:igrj16318-igrj17544} Optical to MIR SEDs of
  IGR~J16318-4848 (left) and IGR~J17544-2619 (right), including data
  from ESO/NTT, VISIR on VLT/UT3 and {\it Spitzer} \citep{rahoui:2008}.
  IGR~J16318-4848 exhibits a MIR excess, interpreted by
  \cite{rahoui:2008} as the signature of a strong stellar outflow
  coming from the sgB[e] companion star \citep{filliatre:2004}.  On the
  other hand, IGR~J17544-2619 is well fitted with only a stellar
  component corresponding to the O9Ib companion star spectral type
  \citep{pellizza:2006}.}
\end{figure}

The multiwavelength observations were performed at the European
Southern Observatory (ESO), using Target of Opportunity (ToO) and
Visitor modes, in 3 domains: optical ($400-800$\,nm) with EMMI, NIR
($1-2.5 \microns$) with SOFI, both instruments at the focus of the
3.5m New Technology Telescope (NTT) at La Silla, and mid-infrared
(MIR, $5-20 \microns$) with the VISIR instrument on Melipal, the 8m
Unit Telescope 3 (UT3) of the Very Large Telescope (VLT) at Paranal
(Chile). With these observations we performed accurate astrometry,
photometry and spectroscopy on a sample constituted of 21 {\it
  INTEGRAL} sources, aiming at identifying their counterparts and the nature of
the companion star, deriving their distance, and finally characterising the
presence and temperature of their circumstellar medium.

\nopagebreak
\begin{figure*}[!ht]
\centering
\includegraphics[height=.34\textheight,angle=-90]{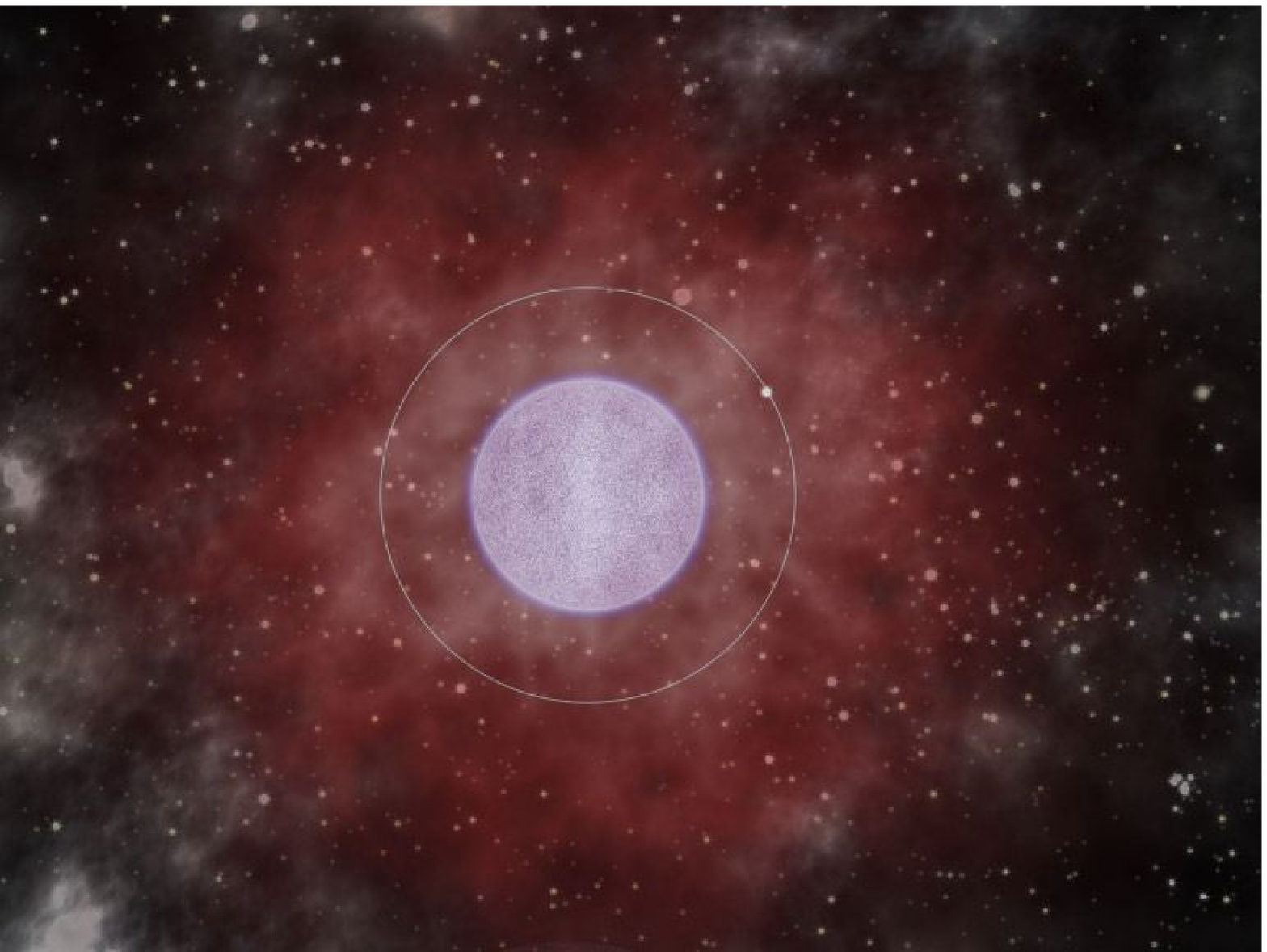}
\includegraphics[height=.34\textheight,angle=-90]{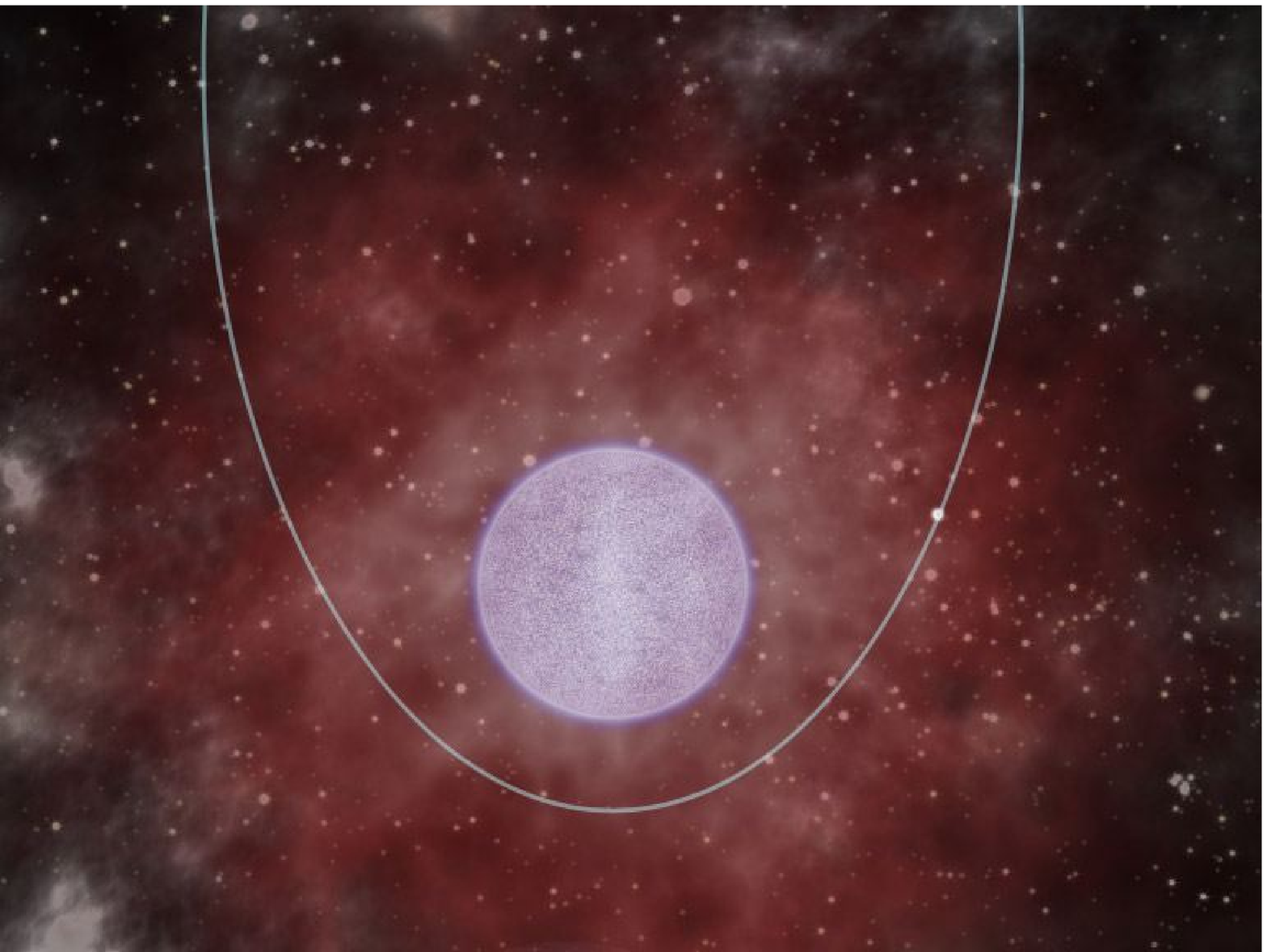}
  \caption[Scenarios illustrating both 2 types of {\it INTEGRAL}
sources]{Scenarios illustrating two possible configurations of {\it
INTEGRAL} sources: a neutron star orbiting around a supergiant
star on a circular orbit (left image); and on an eccentric orbit
(right image), accreting from the clumpy stellar wind of the
supergiant.  The accretion of matter is persistent in the case of the
obscured sources, as in the left image, where the compact object
orbits inside the cocoon of dust enshrouding the whole system. On the
other hand, the accretion is intermittent in the case of SFXTs, which
might correspond to a compact object on an eccentric orbit, as in the
right image.  A 3D animation of these sources is available on the
website: \\
{\em http://www.aim.univ-paris7.fr/CHATY/Research/hidden.html}}
  \label{figure:obscured-sfxt}
\end{figure*}
\nopagebreak
     \subsection{IGR~J16318-4848: Extreme among the Obscured High-Energy
Sources}

IGR~J16318-4848 was the first source discovered by IBIS/ISGRI on
{\it INTEGRAL} on 29 January 2003 \citep{courvoisier:2003}.
{\it XMM-Newton} observations revealed an unusually high level of absorption: 
$\nh \sim 2 \times 10^{24} \cmmoinsdeux$ \citep{matt:2003}.  The accurate
localisation by {\it XMM-Newton} allowed \cite{filliatre:2004} to
rapidly trigger ToO photometric and spectroscopic observations in
optical/NIR, leading to the discovery of the optical counterpart
and to the confirmation of the NIR one \citep{walter:2003}.
The extremely bright NIR source 
(Ks\,$=7.20 \mags$) exhibits an unusually strong intrinsic
absorption in the optical ($A_v = 17.4 \mags$), much stronger than
the absorption along the line of sight ($A_v = 11.4 \mags$), but
still 100 times lower than the absorption in X-rays.  This led
\cite{filliatre:2004} to suggest that the material absorbing in
X-rays was concentrated around the compact object, while the material
absorbing in optical/NIR was enshrouding the whole system.  The
NIR spectroscopy 
revealed an unusual
spectrum, with many strong emission lines, originating from a highly
complex and stratified circumstellar environment of various densities
and temperatures, suggesting the presence of an envelope and strong
stellar outflow responsible for the absorption. Only luminous
early-type stars such as supergiant sgB[e] show such extreme
environments, and \cite{filliatre:2004} concluded that IGR~J16318-4848
was an unusual HMXB.
By combining these optical and NIR data with MIR observations, and
fitting these observations with a model of a sgB[e] companion star,
\cite{rahoui:2008} showed that IGR~J16318-4848 exhibits a MIR
excess (see Figure \ref{figure:igrj16318-igrj17544}, left panel), that they
interpreted as being due to the strong stellar outflow emanating from
the sgB[e] companion star.  They found that the companion star had a
temperature of T\,$=22200$\,K and radius R$_{\star} = 20.4 R_{\odot}$, and
an extra component of temperature T $=1100$\,K and radius R\,$= 10
R_{\star}$, with A$_v = 17.6 \mags$. 
%
%
By taking a typical orbital period of 10 days and a
mass of the companion star of $20 \Msol$, we obtain an orbital
separation of $50 \Rsol$, smaller than the
extension of the extra
component, suggesting that this component enshrouds the whole binary system,
as would do a cocoon of gas/dust (see Figure \ref{figure:obscured-sfxt},
left panel).
In summary, IGR~J16318-4848 is an HMXB system, located at a distance
between 1 and 6 kpc, hosting a compact object (probably a neutron star)
and a sgB[e] star (it is therefore the second HMXB with a sgB[e] star,
after CI Cam; \citeauthor{clark:1999} \citeyear{clark:1999}). The most
striking facts are (i) the compact object seems to be surrounded by
absorbing material and (ii) the whole binary system seems to be
surrounded by a dense and absorbing circumstellar material envelope or
cocoon, made of cold gas and/or dust. This source exhibits such extreme
characteristics that it might not be fully representative of the other
obscured sources.

     \subsection{IGR~J17544-2619: Archetype of the Supergiant Fast
X-ray Transients}

SFXTs constitute a new class of sources identified among the recently
discovered {\it INTEGRAL} sources, exhibiting these common
characteristics: rapid outbursts lasting only hours, a faint quiescent
emission, high energy spectra requiring a BH or NS accretor, and O/B
supergiant companion stars. IGR~J17544-2619, a bright recurrent
transient X-ray source discovered by {\it INTEGRAL} on 17 September
2003 \citep{sunyaev:2003b}, seems to be their archetype. Observations
with {\it XMM-Newton} have shown that it exhibits a very hard X-ray
spectrum, and a relatively low intrinsic absorption ($10^{22}
\cmmoinsdeux$, \citeauthor{gonzalez-riestra:2004}
\citeyear{gonzalez-riestra:2004}).  Its bursts last for hours, and
in between bursts it exhibits long quiescent periods, which can reach
more than 70 days (Zurita Heras \& Chaty, in prep.). The compact
object is probably a neutron star \citep{intzand:2005}.
\cite{pellizza:2006} managed to get optical/NIR ToO observations only
one day after the discovery of this source. They identified a likely
counterpart inside the {\it XMM-Newton} error circle, confirmed by
an accurate localization from {\it Chandra}.  Spectroscopy showed that the
companion star was a blue supergiant of spectral type O9Ib, with a
mass of $25-28 M_{\odot}$ and temperature of T $\sim 31000$ K; the
system is therefore an HMXB \citep{pellizza:2006}.
\cite{rahoui:2008} combined optical, NIR and MIR observations and
showed that they could accurately fit the observations with a model of
an O9Ib star, with temperature T~$=31000$~K and radius R$_{\star} = 21.9
R_{\odot}$. They derived an absorption A$_v = 6.1 \mags$ and a
distance D~$=3.6$~kpc.  The source does not exhibit any MIR excess
(see Figure \ref{figure:igrj16318-igrj17544}, right panel,
 \citeauthor{rahoui:2008} \citeyear{rahoui:2008}).
In summary, IGR~J17544-2619 is an HMXB at a distance of $\sim$3.6~kpc,
constituted of an O9Ib supergiant, with a mild stellar wind and a
compact object which is probably a neutron star, without any MIR excess.

\begin{sidewaystable*}
  \begin{center}
    \caption{Results on the sample of {\it INTEGRAL} sources; more
details are given in \cite{chaty:2008}. We indicate respectively the
name of the sources, the region of the Galaxy in the direction 
which they are located, their spin and orbital period,
the interstellar, optical-IR, and X-ray derived column density
respectively (in units of $10^{22}\cmmoinsdeux$), their spectral type,
nature and reference.
Type abbreviations: 
AGN = Active Galactic Nucleus,
B = Burster,
BHC = Black Hole Candidate,
CV = Cataclysmic Variable,
D = Dipping source,
H = High Mass X-ray Binary system, 
IP = Intermediate polar, 
L = Low Mass X-ray Binary,
O = Obscured source,
P = Persistent source,
S = Supergiant Fast X-ray Transient,
T: Transient source,
XP: X-ray Pulsar.
Reference are:
c: \cite{chaty:2008},
co: \cite{combi:2006},
f: \cite{filliatre:2004},
h: \cite{hannikainen:2007},
m1: \cite{masetti:2004}
m2: \cite{masetti:2006}
n1: \cite{negueruela:2005},
n2: \cite{negueruela:2006a},
n3: \cite{nespoli:2007},
p: \cite{pellizza:2006},
t: \cite{tomsick:2006a}.
}
\label{table:results}
\vspace{1em}
    \renewcommand{\arraystretch}{1.2}
    \begin{tabular}{cccccccccc} 
\hline
Source & Reg & P$_s$(s) & P$_o$(d) & $Nh_{is}$ & $Nh_{IR}$ & $Nh_X$ & SpT & Type & Ref \\
\hline
IGR J16167-4957 & No & & & 2.2 & 0.23 & 0.5 & A0 & CV/IP & t,m2 \\
\hline 
IGR J16195-4945 & No & & & 2.18 & 2.9 & 7 & OB & H?/S?/O & t \\
\hline
IGR J16207-5129 & No & & & 1.73 & 2.0 & 3.7 & BOI & H/O & t,m2 \\
\hline
IGR J16318-4848 & No & & & 2.06 & 3.3 & 200 & sgB[e] & H/O/P & f \\
\hline
IGR J16320-4751 & No & 1250 & 8.96(1) & 2.14 & 6.6 & 21 & sgOB & H/XP/T/O & c \\
\hline
IGR J16358-4726 & No & 5880 & & 2.20 & 3.3 & 33 & sgB[e]? & H/XP/T/O & c \\
\hline
IGR J16393-4643 & No & 912 & 3.6875(6) & 2.19 & 2.19 & 24.98 & BIV-V? & H/XP/T & c \\
\hline 
IGR J16418-4532 & No & 1246 & 3.753(4) & 1.88 & 2.7 & 10 & sgOB? & H/XP/S & c \\
\hline
IGR J16465-4507& No & 228 & & 2.12 & 1.1 & 60 & B0.5I & H/S & n1 \\
\hline
IGR J16479-4514 & No & & & 2.14 & 3.4 & 7.7 & sgOB & H/S? & c \\
\hline
IGR J16558-5203 & - & - & - & - & - & - & Sey1.2 & AGN & m2 \\
\hline 
IGR J17091-3624 & GC & & & 0.77 & 1.03 & 1.0 & L & L/BHC & c \\
\hline 
IGR J17195-4100 & GC & & & 0.77 & & 0.08 & & CV/IP & t,m2 \\
\hline 
IGR J17252-3616 & GC & 413 & 9.74(4) & 1.56 & 3.8 & 15 & sgOB & H/XP/O & c \\
\hline
IGR J17391-3021 & GC & & & 1.37 & 1.7 & 29.98 & O8Iab(f) & H/S/O & n2 \\
\hline
IGR J17544-2619 & GC & & $\geq$70? & 1.44 & 1.1 & 1.4 & O9Ib & S & p \\ 
\hline
IGR J17597-2201 & GC & & & 1.17 & 2.84 & 4.50 & L & L/B/D/P & c \\
\hline
IGR J18027-1455 & - & - & - & - & - & - & Sey1 & AGN & m1,co \\
\hline 
IGR J18027-2016 & GC & 139 & 4.5696(9) & 1.04 & 1.53 & 9.05 & sgOB & H/XP/T & c \\
\hline
IGR J18483-0311 & GC & 21.05 & 18.55 & 1.62 & 2.45 & 27.69 & sgOB? & H/XP & c \\
\hline 
IGR J19140+0951 &    & & 13.558(4) & 1.68 & 2.9 & 6 & sgB0.5I & H/O & n3,h \\
\hline
\end{tabular}
  \end{center}
\end{sidewaystable*}

     \section{General Results on {\it INTEGRAL} Sources and Discussion} \label{IGRs}

To better characterize this population, \cite{chaty:2008} and
\cite{rahoui:2008} studied a sample of 21 {\it INTEGRAL} sources
belonging to both classes described above.
Some results are reported in Table \ref{table:results}. The optical/NIR
study, through accurate astrometry, photometry, and spectroscopy,
allowed \cite{chaty:2008} to identify 
the counterparts, and to show that most of these systems are HMXBs
containing massive and luminous early-type companion stars. By
combining MIR photometry, and fitting their optical--MIR spectral
energy distributions, \cite{rahoui:2008} showed that (i) most of these
sources exhibit an intrinsic absorption and (ii) three of them exhibit
a MIR excess, which they suggest is due to the presence of a cocoon
of dust and/or cold gas enshrouding the whole binary system (see also
\citeauthor{chaty:2006c} \citeyear{chaty:2006c}).
Nearly all the {\it INTEGRAL} HMXBs for which both spin and orbital
periods have been measured are located in the upper part of the Corbet
diagram \citep{corbet:1986}.  They are wind accretors, typical of
supergiant HMXBs, and X-ray pulsars exhibiting longer pulsation
periods and higher absorption (by a factor $\sim4$) as compared to the
average of previously known HMXBs \citep{bodaghee:2007}. This extra
absorption might be due to the presence of a cocoon of dust/cold gas 
enshrouding the whole binary system in the case of the
obscured sources.  The intrinsic properties of the supergiant companion star 
could therefore explain some properties of these
sources.  However, differences exist between obscured
sources and SFXTs, which might be explained by the geometry of the
binary systems, and/or the extension of the wind/cocoon enshrouding
either the companion star or the whole system.  Indeed, obscured
sources are naturally explained by a compact object orbiting inside a
cocoon of dust and/or cold gas, while the fast X-ray behaviour of
SFXTs needs a clumpy stellar wind environment, to account for fast and
transient accretion phenomena (see Figure \ref{figure:obscured-sfxt},
left and right panels respectively, and \citeauthor{chaty:2006c}
\citeyear{chaty:2006c}).
%
These results show the existence in our Galaxy of a dominant
population of a previously rare class of high-energy binary systems:
supergiant HMXBs, some exhibiting a high intrinsic absorption
(\citeauthor{chaty:2008} \citeyear{chaty:2008};
\citeauthor{rahoui:2008} \citeyear{rahoui:2008}).  A careful study of
this population, recently revealed by {\it INTEGRAL}, will provide a
better understanding of the formation and evolution of short-living
HMXBs.  Furthermore, stellar population models will henceforth have to
take these objects into account, to assess a realistic number of
high-energy binary systems in our Galaxy. Our final word is that only
a multiwavelength study reveal the nature of these
obscured high-energy sources.

\acknowledgements
SC would like to thank the organisers for their invitation to report on
these exciting results about newly discovered {\it INTEGRAL} sources, 
and for organising an interesting workshop, in a nice place,
fruitful for scientific discussions and new ideas to arise. 


\begin{thebibliography}{}

\bibitem[{Bird} et~al., 2007]{bird:2007}
{Bird}, A.~J., {Malizia}, A., {Bazzano}, A., {Barlow}, E.~J., {Bassani}, L.,
  {Hill}, A.~B., {B{\'e}langer}, G., {Capitanio}, F., {Clark}, D.~J., {Dean},
  A.~J., {Fiocchi}, M., {G{\"o}tz}, D., {Lebrun}, F., {Molina}, M., {Produit},
  N., {Renaud}, M., {Sguera}, V., {Stephen}, J.~B., {Terrier}, R., {Ubertini},
  P., {Walter}, R., {Winkler}, C., and {Zurita}, J. (2007).
\newblock {The Third IBIS/ISGRI Soft Gamma-Ray Survey Catalog}.
\newblock {\em \apjs}, 170:175--186.

\bibitem[{Bodaghee} et~al., 2007]{bodaghee:2007}
{Bodaghee}, A., {Courvoisier}, T.~J.-L., {Rodriguez}, J., {Beckmann}, V.,
  {Produit}, N., {Hannikainen}, D., {Kuulkers}, E., {Willis}, D.~R., and
  {Wendt}, G. (2007).
\newblock {A description of sources detected by INTEGRAL during the first 4
  years of observations}.
\newblock {\em \aap}, 467:585--596.

\bibitem[{Chaty} \& {Filliatre}, 2005]{chaty:2005a}
{Chaty}, S. \& {Filliatre}, P. (2005).
\newblock {Revealing the Nature of the Obscured High Mass X-ray Binary IGR
  J16318-4848}.
\newblock {\em \apss}, 297:235--244.

\bibitem[{Chaty} \& {Rahoui}, 2006]{chaty:2006c}
{Chaty}, S. \& {Rahoui}, F. (2006).
\newblock {Optical to Mid-infrared observations revealing the most obscured
  high-energy sources of the Galaxy}.
\newblock In {\em Procs. of 6th INTEGRAL workshop, Moscow, Russia}.
\newblock in press (astro-ph/0609474).

\bibitem[{Chaty} et~al., 2008]{chaty:2008}
{Chaty}, S., {Rahoui}, F., {Foellmi}, C., {Rodriguez}, J., {Tomsick}, J.~A.,
  and {Walter}, R. (2008).
\newblock The most obscured high-energy sources of the galaxy brought to light
  by optical/infrared observations.
\newblock {\em \aap}.
\newblock in press.

\bibitem[{Clark} et~al., 1999]{clark:1999}
{Clark}, J.~S., {Steele}, I.~A., {Fender}, R.~P., and {Coe}, M.~J. (1999).
\newblock {Near IR spectroscopy of candidate B[e]/X-ray binaries}.
\newblock {\em \aap}, 348:888--896.

\bibitem[{Combi} et~al., 2006]{combi:2006}
{Combi}, J.~A., {Rib{\'o}}, M., {Mart{\'{\i}}}, J., and {Chaty}, S. (2006).
\newblock {Multi-wavelength properties of the high-energy bright Seyfert 1
  galaxy IGR J18027-1455}.
\newblock {\em \aap}, 458:761--766.

\bibitem[{Corbet}, 1986]{corbet:1986}
{Corbet}, R.~H.~D. (1986).
\newblock {The three types of high-mass X-ray pulsator}.
\newblock {\em \mnras}, 220:1047--1056.

\bibitem[{Courvoisier} et~al., 2003]{courvoisier:2003}
{Courvoisier}, T.~J.-L., {Walter}, R., {Rodriguez}, J., {Bouchet}, L., and
  {Lutovinov}, A.~A. (2003).
\newblock {Igr J16318-4848}.
\newblock {\em IAU Circ.}, 8063:3--+.

\bibitem[{Filliatre} \& {Chaty}, 2004]{filliatre:2004}
{Filliatre}, P. \& {Chaty}, S. (2004).
\newblock {The Optical/Near-Infrared Counterpart of the INTEGRAL Obscured
  Source IGR J16318-4848: An sgB[e] in a High-Mass X-Ray Binary?}
\newblock {\em \apj}, 616:469--484.

\bibitem[{Gonz{\'a}lez-Riestra} et~al., 2004]{gonzalez-riestra:2004}
{Gonz{\'a}lez-Riestra}, R., {Oosterbroek}, T., {Kuulkers}, E., {Orr}, A., and
  {Parmar}, A.~N. (2004).
\newblock {XMM-Newton observations of the INTEGRAL X-ray transient IGR
  J17544-2619}.
\newblock {\em \aap}, 420:589--594.

\bibitem[{Hannikainen} et~al., 2007]{hannikainen:2007}
{Hannikainen}, D.~C., {Rawlings}, M.~G., {Muhli}, P., {Vilhu}, O., {Schultz},
  J., and {Rodriguez}, J. (2007).
\newblock {The nature of the infrared counterpart of IGR J19140+0951}.
\newblock {\em \mnras}, 380:665--668.

\bibitem[{in't Zand}, 2005]{intzand:2005}
{in't Zand}, J.~J.~M. (2005).
\newblock {Chandra observation of the fast X-ray transient IGR J17544-2619:
  evidence for a neutron star?}
\newblock {\em \aap}, 441:L1--L4.

\bibitem[{Masetti} et~al., 2006]{masetti:2006}
{Masetti}, N., {Morelli}, L., {Palazzi}, E., {Galaz}, G., {Bassani}, L.,
  {Bazzano}, A., {Bird}, A.~J., {Dean}, A.~J., {Israel}, G.~L., {Landi}, R.,
  {Malizia}, A., {Minniti}, D., {Schiavone}, F., {Stephen}, J.~B., {Ubertini},
  P., and {Walter}, R. (2006).
\newblock {Unveiling the nature of INTEGRAL objects through optical
  spectroscopy. V. Identification and properties of 21 southern hard X-ray
  sources}.
\newblock {\em \aap}, 459:21--30.

\bibitem[{Masetti} et~al., 2004]{masetti:2004}
{Masetti}, N., {Palazzi}, E., {Bassani}, L., {Malizia}, A., and {Stephen},
  J.~B. (2004).
\newblock {Unveiling the nature of three INTEGRAL sources through optical
  spectroscopy}.
\newblock {\em \aap}, 426:L41.

\bibitem[{Matt} \& {Guainazzi}, 2003]{matt:2003}
{Matt}, G. \& {Guainazzi}, M. (2003).
\newblock {The properties of the absorbing and line-emitting material in IGR
  J16318 - 4848}.
\newblock {\em \mnras}, 341:L13--L17.

\bibitem[{Negueruela} et~al., 2005]{negueruela:2005}
{Negueruela}, I., {Smith}, D.~M., and {Chaty}, S. (2005).
\newblock {Optical counterpart to IGR J16465-4507}.
\newblock {\em The Astronomer's Telegram}, 429:1--+.

\bibitem[{Negueruela} et~al., 2006]{negueruela:2006a}
{Negueruela}, I., {Smith}, D.~M., {Reig}, P., {Chaty}, S., and {Torrej{\'o}n},
  J.~M. (2006).
\newblock {Supergiant Fast X-ray Transients: a new class of high mass X-ray
  binaries unveiled by INTEGRAL}.
\newblock In {Wilson}, A., editor, {\em ESA Special Publication}, volume 604 of
  {\em ESA Special Publication}, pages 165--170.

\bibitem[{Nespoli} et~al., 2007]{nespoli:2007}
{Nespoli}, E., {Fabregat}, J., and {Mennickent}, R. (2007).
\newblock {K-band spectroscopy of AX J1841.0-0536 and IGR J19140+0951.}
\newblock {\em The Astronomer's Telegram}, 982:1--+.

\bibitem[{Pellizza} et~al., 2006]{pellizza:2006}
{Pellizza}, L.~J., {Chaty}, S., and {Negueruela}, I. (2006).
\newblock {IGR J17544-2619: a new supergiant fast X-ray transient revealed by
  optical/infrared observations}.
\newblock {\em \aap}, 455:653--658.

\bibitem[{Rahoui} et~al., 2008]{rahoui:2008}
{Rahoui}, F., {Chaty}, S., {Lagage}, P.-O., and {Pantin}, E. (2008).
\newblock The most obscured high-energy sources of the galaxy brought to light
  by optical/infrared observations.
\newblock {\em \aap}.
\newblock in press.

\bibitem[{Sguera} et~al., 2005]{sguera:2005}
{Sguera}, V., {Barlow}, E.~J., {Bird}, A.~J., {Clark}, D.~J., {Dean}, A.~J.,
  {Hill}, A.~B., {Moran}, L., {Shaw}, S.~E., {Willis}, D.~R., {Bazzano}, A.,
  {Ubertini}, P., and {Malizia}, A. (2005).
\newblock {INTEGRAL observations of recurrent fast X-ray transient sources}.
\newblock {\em \aap}, 444:221--231.

\bibitem[{Sunyaev} et~al., 2003]{sunyaev:2003b}
{Sunyaev}, R.~A., {Grebenev}, S.~A., {Lutovinov}, A.~A., {Rodriguez}, J.,
  {Mereghetti}, S., {Gotz}, D., and {Courvoisier}, T. (2003).
\newblock {New source IGR J17544-2619 discovered with INTEGRAL}.
\newblock {\em The Astronomer's Telegram}, 190:1--+.

\bibitem[{Tomsick} et~al., 2006]{tomsick:2006a}
{Tomsick}, J.~A., {Chaty}, S., {Rodriguez}, J., {Foschini}, L., {Walter}, R.,
  and {Kaaret}, P. (2006).
\newblock {Identifications of Four INTEGRAL Sources in the Galactic Plane via
  Chandra Localizations}.
\newblock {\em \apj}, 647:1309--1322.

\bibitem[{Walter} et~al., 2003]{walter:2003}
{Walter}, R., {Rodriguez}, J., {Foschini}, L., {de Plaa}, J., {Corbel}, S.,
  {Courvoisier}, T.~J.-L., {den Hartog}, P.~R., {Lebrun}, F., {Parmar}, A.~N.,
  {Tomsick}, J.~A., and {Ubertini}, P. (2003).
\newblock {INTEGRAL discovery of a bright highly obscured galactic X-ray binary
  source IGR J16318-4848}.
\newblock {\em \aap}, 411:L427--L432.

\end{thebibliography}

\end{document}